\documentclass[prd,eqsecnum,twocolumn,amsfonts,showpacs]{revtex4}
\newcommand{\beq}{\begin{equation}}
\newcommand{\eeq}{\end{equation}}
\newcommand{\beqs}{\begin{eqnarray}}
\newcommand{\eeqs}{\end{eqnarray}}
\newcommand{\lsim}{\mathrel{\raisebox{-
.6ex}{$\stackrel{\textstyle<}{\sim}$}}}
\newcommand{\gsim}{\mathrel{\raisebox{-
.6ex}{$\stackrel{\textstyle>}{\sim}$}}}

\begin{document}

\title{Gluon-Glueball Duality and Glueball Searches} 

\author{Shmuel Nussinov$^{a,b}$}

\author{Robert Shrock$^c$}

\affiliation{(a)
School of Physics and Astronomy, Tel Aviv University, Tel Aviv, Israel}

\affiliation{(b)
Schmid College of Science, Chapman University, Orange, CA 92866} 

\affiliation{(c)
C.N. Yang Institute for Theoretical Physics, Stony Brook University,
Stony Brook, NY 11794}

\begin{abstract}

We discuss a notion of gluon-glueball duality analogous to quark-hadron
duality. We apply this idea to the radiative decay of heavy orthoquarkonium, $Q
\bar Q \to \gamma g g$, which has been used to search for glueballs.  The
duality is first introduced in two simplified contexts: (i) a hypothetical
version of QCD without any light quarks and (ii) QCD in the large-$N_c$ limit.
We then discuss how an approximate form of this duality could hold in real QCD,
based on a hierarchy of time scales in the temporal evolution of the $gg$
subsystem in radiative orthoquarkonium decay.  We apply this notion of
gluon-glueball duality to suggest a method that could be useful in experimental
searches for glueballs.

\end{abstract}


\maketitle

\section{Introduction}

Quantum chromodynamics (QCD) is very successful theory describing quark and
gluon interactions. There are ample observations of gluon jets in high-energy
collider data, and lattice QCD calculations of the pure gluonic sector of the
theory have yielded a detailed spectrum of (color-singlet) bound states of
gluons, commonly called glueballs \cite{besreview}-\cite{pdg}. The lightest of
these can be modelled as $gg$ states, where $g$ denotes a gluon; these include
a state with $S=0$, $L=0$, and $J^{PC}=0^{++}$, and a heavier state with $S=2$,
$L=0$, and $J^{PC}=2^{++}$.  Radial excitations, states with angular momentum
$L \ge 1$, and $ggg$ states also appear in the spectrum.  Over the years there
have been numerous experimental searches for glueballs.  It was pointed out
early on that a promising method is to use the radiative decay of a heavy $Q
\bar Q$ orthoquarkonium state \cite{brodsky78,koller78}.  At the level of
elementary constituents, this decay is $Q \bar Q \to \gamma g g$, so that when
the two gluons are emitted with an invariant mass close to that of a glueball,
they have substantial probability to bind to form this state.  Other production
channels have also been used.  At present, there are strong indications for
hadrons with large gluonic components, although there are still is no consensus
concerning the details of the mixing of $q \bar q$ and gluonic components to
form various physical mass eigenstates \cite{besreview}-\cite{pdg}.

   In this paper we examine the temporal evolution of glueball production in
radiative orthoquarkonium decay. We use the fact that glueballs have a smaller
density of states than $q \bar q$ mesons, as a function of mass, in conjunction
with the Heisenberg uncertainty principle, to infer that one can generically
measure the formation of a glueball sooner than the formation of a $q \bar q$
meson.  On the basis of this observation, we propose a notion of gluon-glueball
duality.  We apply this to comment on current experimental searches for
glueballs and to suggest a method that could be useful for these searches.  An
outline of the paper is as follows. In Sect. II we review quark-hadron
duality. In Sects. III and IV we give some background on glueball properties
and searches.  In Sect. V we introduce the notion of gluon-glueball duality in
two simplified contexts, and in Sect. VI we discuss it in full QCD.  We point
out that in studying the production and decay of glueballs, it is useful to
analyze the temporal evolution of the $gg$ subsystem as it is produced, binds
to form a proto-glueball, mixes with $q \bar q$ components, and finally decays.
Section VII suggests some future lattice gauge measurements that are relevant
to gluon-glueball duality, while in Sect. VIII we apply our observations to
experimental searches for glueballs.

\section{Quark-Hadron Duality}

We first give some background on ideas of duality in hadronic physics.  The
reader who is familiar with this material can skip this section and proceed
directly to our new observations in Sects. V and VI.  The idea of quark-hadron
duality in several related forms \cite{hr,bg} dates back to the early period
in the development of the quark-parton model. In the Bloom-Gilman form
\cite{bg}, it states, roughly speaking, that in a reaction such as a electron
scattering off a nucleon, the sum of the cross sections for the full set of
exclusive hadronic final states $X_h$ that are kinematically accessible at a
given center-of-mass energy $E_{cm}=\sqrt{s}$ is equivalent to the cross
section for the elementary reaction $e + q \to e + q$ involving the quarks in
the nucleon.  A similar duality relation applies to charged-current
neutrino reactions such as $\nu_\mu + N \to \mu + X_h$.  Let us denote the
four-momenta of the incident and scattered leptons as $\ell_1$ and $\ell_2$,
with $\ell_1 - \ell_2 = q$, $(\ell_1^0)_{lab}=E$, $(\ell_2^0)_{lab}=E'$, and
the four-momenta of the target nucleon and final hadronic state as $p$ and
$p_X$. We further recall the standard Bjorken variables $x=-q^2/(2 q \cdot p)$
and $y=q \cdot p/\ell_1 \cdot p = (E-E')/E$. Then this duality is the statement
that
\beq
\sum_{X_h} \sigma(\nu_\mu + N \to \mu + X_h) \sim \int_0^1 dx \, \int_0^1 dy
 \, \frac{d\sigma}{dxdy}(\nu_\mu + f \to \mu + f') \ , 
\label{slacdis}
\eeq
where $f$ denotes all of the charge $-1/3$ quarks (and charge $-2/3$
antiquarks) that can participate in this reaction.  At a fundamental level,
this duality is justified by the asymptotic freedom of QCD \cite{af}.  In the
deep inelastic scattering of an electron or neutrino off of a nucleon $N$, the
hadronic part of the cross section involves the tensor
\beqs
W_{\mu\nu}(q,p) & = & \frac{1}{2}\sum_X \langle N | J_\mu |X\rangle 
                                    \langle X | J_\nu^\dagger | N \rangle  
(2\pi)^3 \delta(p+q-p_X) \cr\cr
                & \propto & \int \frac{d^4z}{2\pi} e^{-iq \cdot z} 
\langle N | J_{\mu}(z) J_\nu(0)^\dagger | N \rangle \ , \cr\cr
& & 
\label{wmunu}
\eeqs
where $X$ denotes a hadronic final state and $J_\mu$ is the respective
electromagnetic or weak (charged or neutral) current.  One then uses the Wilson
operator product expansion to express the bilocal product of currents in terms
of a sum of local operators, applicable near to the light cone $z^2 \to 0$, as
enforced by the kinematic conditions $-q^2 >> \Lambda_{QCD}^2$ and $q \cdot p
>> \Lambda_{QCD}^2$, where $\Lambda_{QCD} \simeq 300$ MeV is the scale where
QCD confines and spontaneously breaks chiral symmetry.  This enables one to
express the deep inelastic scattering off the nucleon in terms of the
scattering off of quarks.  The asymptotic freedom of QCD has the consequence
that these quarks are quasi-free when probed at short distances.  Similarly,
away from particle thresholds, one can calculate the total cross section for
$e^+e^- \to {\rm hadrons}$ at center-of-mass energy $\sqrt{s}$ in terms of the
cross section for $e^+e^- \to q \bar q$, where $2m_q \lsim \sqrt{s}$.
One can consider the cross section for $e^+e^- \to {\rm hadrons}$,
smeared over resonances, to be equivalent to the elementary reaction $e^+e^-
\to q \bar q$, summed over the kinematically accessible quarks
\cite{ap}-\cite{smear}
\beqs
\sum_{X_h} \sigma(e^+ e^- \to X_h) \sim \sum_q \sigma(e^+ e^- \to q \bar q) \ .
\label{epluseminus}
\eeqs

  In the full QCD theory, the notion of quark-hadron duality is naturally
generalized to parton-hadron duality, where the partons include both quarks and
gluons, and the hadrons are understood to include not only $qqq$ baryons and $q
\bar q$ mesons, but also hadronic mass eigenstates that are linear combinations
of $q \bar q$ and $gg$, $ggg$, etc.  Possible exotic color-singlet hadrons such
as, in the bosonic sector, $q \bar q q \bar q$ and $q \bar q g$ can, in
principle, also be included in this set of physical states.  In one sense, this
duality amounts to the statements that (i) there is a complete orthonormal
basis of perturbative quark and gluon states forming the Fock space of
perturbative QCD, and there is a complete orthonormal basis of physical
color-singlet hadronic mass eigenstates forming another Fock space; and (ii),
given the asymptotic freedom of QCD, the cross section for an inclusive
reaction involving the contributions of many exclusive physical channels with
smearing over resonances as appropriate, can be expressed in terms of the
corresponding cross section in terms of the elementary partonic degrees of
freedom. In another sense, one can think of it as somewhat analogous to a
Mittag-Leffler expansion, in which a function is written as a sum over its
poles.  In this context, one may recall that the Mittag-Leffler expansion of
the Euler beta function forms part of the mathematical basis of the $s$-$t$
duality in the Veneziano and Virasoro amplitudes in hadronic string theory
\cite{dhs}-\cite{gsw}.

A specific $\bar q q \leftrightarrow $ meson duality (and the analogous $gg
\leftrightarrow $ glueball duality to be introduced next) is particularly
useful. This is especially the case if one considers the large-$N_c$ limit of
QCD \cite{thooft,witten,jaffe}.  For large $N_c$, baryons become very heavy,
and the kinematically accessible hadronic states $X_h$ directly produced in
$e^+e^-$ annhilation are $\bar qq$ mesons. Since the decay rate of such a meson
or glueball vanishes in the large-$N_c$ limit, meson resonances are narrow in
this limit. The energy integral in Eq. (\ref{epluseminus}) then becomes
essentially a summation over the contributions of these resonances.  

Mesons and baryons are observed to lie on approximately linear Regge
trajectories of the form
\beq
\alpha(m^2) = \alpha_0 + \alpha' m^2
\label{regge}
\eeq
with respective intercepts $\alpha_0$ and a common Regge slope $\alpha' = 0.9$
GeV$^{-1}$.  Physical meson states occur where the angular momentum
$\alpha(m^2)$ is equal to a non-negative integer.  This behavior was originally
motivated by analysis of potential scattering and was elegantly explained by
hadronic string theory (the dual resonance model), according to which a meson
is a mass eigenstate of an open string.  It is believed (although it has not
been proved) that the large-$N_c$ limit of SU($N_c$) QCD reproduces features of
a hadronic string theory.  In the hadronic string model, the string tension
$\sigma=1/(2\pi \alpha')$, so that $\sqrt{\sigma} \simeq 0.42$ GeV.
Physically, this string tension represents the energy per unit length of the
chromoelectric flux tube between the $q$ and $\bar q$ forming the meson.  An
example of a Regge trajectory is that for the $S=1$, $I=1$ (isovector) mesons,
which includes $\rho(770)$, $a_2(1320)$, $\rho_3(1690)$, and $a_4(2040)$, with
increasing values of $J$ indicated as subscripts (where $\vec J = \vec L + \vec
S$).  The radial excitations $\rho' = \rho(1450)$, $\rho'' = \rho(1700)$,
etc. are on so-called daughter trajectories, forming a horizontal line in the
plane with horizontal and vertical axes corresponding to $s=m^2$ and $J$,
respectively.

A feature predicted by hadronic string theory (predating QCD) and consistent
with data is that the density of $q \bar q$ meson states as a function of mass
$m$ grows rapidly with $m$.  This is also the case for a specific flavor state
such as $\bar u d$ and specific values of $J$, parity and charge conjugation
quantum numbers, such as $J^{PC}=1^{--}$. Let us denote the density of meson
states, i.e., the number of states at a given mass $m$, counting those on the
leading and daughter meson trajectories, as
\beq
n(m)_M \equiv \frac{dn_M(m)}{dm} \ , 
\label{nmdef}
\eeq
where $M$ stands for ``meson''. For the (bosonic) string in $d$ spacetime
dimensions, the meson density of states $n(m)_M$, grows exponentially fast for
$m^2 >> (\alpha')^{-1}$ \cite{nm,hag}:
\beq
n(m)_M \sim m^{-(d+1)/4} \, \exp \Big [ \pi m \sqrt{(2/3)(d-2)\alpha'} \ \Big ]
\label{nm}
\eeq
where $d$ is the spacetime dimension.  Hence, at sufficiently high mass, these
resonances overlap. Indeed, even before one takes account of this asymptotic
exponential growth in the density of states, the hadronic string model already
implies that they will overlap, because on the leading Regge trajectory,
Eq. (\ref{regge}) shows that two successive meson states with the same
$J^{PC}$, that differ by two units of $L$ and $J$, satisfy $\Delta J =
2=\alpha'(m_{L+2}^2-m_L^2)$, so that
\beq
m_{L+2} - m_L = \frac{2}{\alpha'(m_{L+2} + m_L)} \ . 
\label{mdif}
\eeq
Hence, as the masses of these states increase, their mass difference decreases,
and eventually becomes less than their widths, so that they overlap.  This
happens when the mass difference $m_{L+2} - m_L$ becomes comparable to to
either of the widths $\Gamma_L$ or $\Gamma_{L+2}$.  For the present
illustrative purposes, we approximate these as being roughly equal, and denote
them as $\Gamma$, which we take to be $\Gamma \sim 0.25$ GeV. Setting
$m_{L+2}-m_L = \Gamma$ and solving, we get
\beq
\frac{m_{L+2}+m_L}{2} \simeq \frac{1}{\alpha' \Gamma} \simeq 4.5 \ {\rm GeV} \
.
\label{mmerge}
\eeq
Thus, as (light-quark) meson masses increase beyond this scale, the states in
their spectrum tend to merge. In the upper end of the mass region of interest
here, from about 1.5 to 3 GeV, the asymptotic condition $m^2 >> (\alpha')^{-1}$
begins to be satisfied, so the formula (\ref{nm}) is relevant. A hadronization
model based on the chromoelectric flux tube between a $q$ and $\bar q$ in
conjunction with a Schwinger mechanism was given in Ref. \cite{fluxtube}.  The
non-Abelian generalization, in which a constant chromoelectric field creates
gluons, was analyzed in terms of relevant invariants in \cite{gg,eigen}.  The
flux-tube mechanism is incorporated in current hadronization computer programs
such as PYTHIA \cite{pythia}. Because of the increasing density of meson states
for masses $m^2 >> (\alpha')^{-1}$, the cross section for $e^- e^+ \to q \bar q
\to$ hadrons then becomes a continous curve which, according to the duality
assumption, coincides with the continous perturbative curve.

An important feature concerns the behavior in the mass region below
approximately 3 GeV.  The asymptotic freedom and precocious scaling properties
of QCD make quark-hadron duality a property that is effectively local in mass
already at masses that are only modestly greater than $\Lambda_{QCD}$.  Thus,
the $\rho$ and $\rho'$ of masses 0.77 and 1.45 GeV largely account for the
contributions in their mass region to finite-energy sum rules \cite{dhs}.  This
is also manifest in Bloom-Gilman duality \cite{bg}.  A difference is that
Dolen-Horn-Schmid duality applies to $2 \to 2$ reactions involving onshell
hadrons, e.g., $\pi^+ \pi^- \to \pi^+ \pi^-$.  Similarly, Bloom-Gilman duality
applies to reactions such as exclusive electroproduction, e.g., $e + p \to e +
p + \pi^0$.

\section{Remarks on Glueball Properties}

In this section we note some properties of glueballs that we will use in our
analysis.  An especially important and relevant property that motivates our new
suggestion is the density of states, but we begin with some basic facts.  Since
the gluons are bosons, Bose statistics implies that the total glueball
wavefunction is symmetric under interchange of any two gluons.  A difference
between $q \bar q$ mesons and glueballs is that although a confined quark picks
up a (gauge-invariant) dynamical, constituent mass of order $\Lambda_{QCD}$,
one cannot ascribe a mass in the same manner to a bound gluon, since this would
violate the color gauge invariance.  This means that while a constituent quark
model can provide a good description of baryons and $q \bar q$ mesons (see,
e.g., \cite{kp} for a recent discussion and references to the literature), one
cannot describe the glueball in quite so simple a manner.  Furthermore, in the
time evolution of an initial gluonic state, the splittings $g \to gg$ can occur
in a manner that is leading in $1/N_c$, in the large-$N_c$ limit. This is
different from the time evolution of a $q \bar q$ state, for which the
transition $q \to q + g$ is suppressed in the large-$N_c$ limit. Thus, here a
physical state denoted as $gg$ strictly refers only to a state whose quantum
numbers are most simply attainable via a (color-singlet) combination of two
gluons.  Keeping this caveat in mind, the lowest-lying glueballs can be
modelled as $gg$ bound states.  For these, in the Clebsch-Gordon decomposition
of the $gg$ SU(3)$_c$ representations $8 \times 8$, the singlet appears as a
symmetric combination.  Hence, the product of the space and spin wavefunctions
must be even under this interchange.  The spin wavefunction involves the
addition of two spin-1 angular momenta.  If the resultant spin of the $gg$
combination is $S=0$ or $S=2$, this spin wavefunction is even, so the relative
angular momentum must also be even, and the ground state is $L=0$.  With
$P=(-1)^L$ and $C=(-1)^{L+S}$ for this combination of two bosons, one thus
expects that the lowest two glueball states have (i) $S=L=J=0$, whence $J^{PC}
= 0^{++}$ and (ii) $L=0$, $S=J=2$, whence $J^{PC}=2^{++}$. The higher-lying
glueball states can involve both nonzero internal angular momenta and radial
excitations.

Estimates of glueball masses and widths have been made on the basis of a number
of different methods \cite{besreview,earlierreviews},
\cite{brodsky78,koller78}, \cite{mt}-\cite{chen06}, Continuum approaches
include the MIT bag model \cite{mit76}-\cite{adscft2}, flux-tube models,
AdS/CFT approaches, and calculations based on the Bethe-Salpeter equation.
Lattice calculations have achieved a rather high level of precision \cite{mt},
\cite{earlyglueball}-\cite{chen06}.  These naturally give the mass of a
particular glueball in terms of the square root of the string tension,
$\sqrt{\sigma} = 0.42$ GeV.  For masses of glueballs in purely gluonic QCD,
recent lattice calculations \cite{lw99,mp99,chen06} yield
\beq
m(0^{++}) \simeq 1.7 \ {\rm GeV} \ , 
\label{m0++}
\eeq
\beq
m(2^{++}) \simeq 2.4 \ {\rm GeV} \ , 
\label{m2++}
\eeq
\beq
m(0^{-+}) \simeq m(0^{++ \ '}) \simeq 2.6 \ {\rm GeV} \ , 
\label{m0-+}
\eeq
and
\beq
m(1^{+-}) \simeq m(2^{-+}) \simeq 3.0 \ {\rm GeV} \ , 
\label{m1+-}
\eeq
up to approximately 3 GeV.  Here, the $0^{++ \ '}$ glueball is a radial
excitation of the $0^{++}$ glueball. Lattice measurements of higher-lying
glueball masses have been made up to roughly 5 GeV \cite{lw99,mp99,chen06}.
Some unquenched calculations have also been reported \cite{hartteper}. 

In the context of the Regge or hadronic string model, glueballs correspond to
closed strings, which have a Regge slope equal to half of the Regge slope for
open strings:
\beq
(\alpha')_{GB} = \frac{\alpha'}{2} \ . 
\label{alphaprimerel}
\eeq
It follows that for $m^2 >> (\alpha')^{-1}$, the density of states for
glueballs (closed strings), $n(m)_{GB} \equiv dn(m)_{GB}/dm$, is exponentially
smaller than the density of states for $q \bar q$ mesons (open strings),
$n(m)_M$. Quantitatively, from Eq. (\ref{nm}) and Eq. (\ref{alphaprimerel}),
one finds that, for $m^2 >> (\alpha')^{-1}$, the ratio of these densities of
states is
\beq
\frac{n(m)_{GB}}{n(m)_M} \sim 2^{-(d+1)/4} \, \exp \Big [ \pi m (\sqrt{2}-1)
\sqrt{(d-2)\alpha'/3} \ \Big ] \ . 
\label{nratio}
\eeq
With $\alpha'=0.9$ GeV$^{-1}$ (and $d=4$), 
\beq
\frac{n(m)_{GB}}{n(m)_M} \sim 0.3 \quad {\rm for} \quad m=2 \ \ {\rm GeV}
\label{ratioat2gev}
\eeq
and
\beq
\frac{n(m)_{GB}}{n(m)_M} \sim 0.1 \quad {\rm for} \quad m=3 \ \ {\rm GeV}
 \ .
\label{ratioat3gev}
\eeq
Thus, Eq. (\ref{nratio}) indicates that the spectrum of low-lying glueball
states is more sparse than that of the isoscalar $q \bar q$ mesons in the mass
region from 1.5 to 3 GeV.  To within the theoretical and experimental
uncertainties, this is consistent with the data: for example, in the mass
region 1.3 to 2 GeV, there are the following scalar $0^{++}$ states
$f_0(1370)$, $f_0(1500)$, $f_0(1710)$, and indications from recent BES data of
an $f_0(1790)$ and $f_0(1810)$ \ \cite{besreview}.  The lattice estimates (to
be discussed next) indicate that in this interval of masses, there is one
$0^{++}$ glueball expected.  In this channel, this gives a ratio of
$n(m)_{GB}/n(m)_M \sim 0.25$.

Estimates have also been made of glueball widths.  In the limit $N_c \to
\infty$ with $g_s^2 N_c$ fixed and finite \cite{thooft,witten}, where $g_s$ is
the SU(3)$_c$ gauge coupling, the width of a glueball vanishes like
\beq
\Gamma_{GB} = \frac{1}{\tau_{GB}} \propto \frac{\Lambda_{QCD}}{N_c^2} \ , 
\label{gammagblargenc}
\eeq
while the width of a $q \bar q$ meson $M$ vanishes like
\beq
\Gamma_M = \frac{1}{\tau_M} \propto \frac{\Lambda_{QCD}}{N_c} \ . 
\label{gammaqqbarlargenc}
\eeq
The relations (\ref{gammagblargenc}) and (\ref{gammaqqbarlargenc}) follow from
direct diagrammatic $1/N_c$ counting.  As expected from the close
correspondence between the large-$N_c$ limit of QCD and the hadronic string
picture, they can also easily be understood in a string picture. The decay of a
$q \bar q$ meson resonance (an open string) takes place via a single cut in the
string (flux-tube), whereas the decay of a glueball requires a first cut to
transform it from the initial closed string to an open string and then a second
cut to produce the two-meson (e.g., $\pi \pi$) final state. With each cut being
suppressed by a $1/N_c$ factor, the results on $\Gamma_M$ and $\Gamma_{GB}$
follow.  Reverting from the large-$N_c$ limit to real QCD, actual estimates of
glueball widths have varied widely, ranging from a few MeV to $O(10^2)$ MeV
\cite{robson,ccfgm81,lw99}.

\section{Previous Searches for Glueballs} 

Here we briefly review results of previous searches for glueballs. There is an
extended literature dealing with search criteria and analysis of data
\cite{besreview,earlierreviews},\cite{brodsky78,koller78},\cite{goldhabergoldman}-\cite{narisonnew}.
One signature is that glueballs would not fit into the standard set of $q \bar
q$ states, including their angular momentum and radial excitations.  Second,
since the gluons carry no electric charge, one expects a small branching ratio
of glueballs into photons. Third, since the gluons carry no flavor, it was
originally expected that the decays of these states should be
flavor-independent, up to phase space considerations.  On the other hand,
however, it has been suggested that for $J=0$ glueballs, there should be
helicity suppression of decays to light-quark hadrons, at least if the decay
amplitude element can be accurately modelled beginning with emission of a
single $q \bar q$ pair \cite{chan05}; if it involves higher initial
multiplicity of (anti)quarks, then this helicity suppression would be reduced
\cite{chao07}. Fourth, some glueball states have exotic values of $J^{PC}$ that
cannot be obtained from $q \bar q$.

Experimental searches for glueballs have been carried out at many laboratories.
Experiments using $e^+e^-$ annihilation include Mark III and the Crystal Ball
at SPEAR, the subsequent Crystal Ball experiment at DORIS, and experiments at
other laboratories, including Orsay, CESR, Novosibirsk, BES, BABAR, and Belle
\cite{besreview}-\cite{pdg}, \cite{mark3}-\cite{belle_res}. We focus first on
the isoscalar, $J^{PC}=0^{++}$ channel, since the lightest pure glueball has
these quantum numbers.  There are three prominent isoscalar, Lorentz scalar
$0^{++}$ meson resonances between about 1.0 and 1.7 GeV, namely the
$f_0(1370)$, $f_0(1500)$, and $f_0(1710)$.  The quark model is only expected to
produce two such states, which would have $S=1$, $L=1$, $J=0$ and be the
analogues of the flavor SU(3) octet and singlet pseudoscalar mesons, $\eta$ and
$\eta'$.  The fact that there are three $f_0$ states in this range is thus one
of several pieces of evidence suggesting that the third may be primarily a
glueball.  The $f_0(1370)$ is quite broad, with $\Gamma \sim 300$ MeV, while
the $f_0(1500)$ and $f_0(1710)$ have widths of roughly 100-140 MeV \cite{pdg}.
More recently, The Beijing $e^+e^-$ collider BES has found evidence for an
$f_0(1790)$ and $f_0(1810)$ \cite{besreview}. Several theoretical fits to these
data have been performed \cite{besreview,earlierreviews}.  The authors of some
of these fits concluded that the lightest glueball forms a primary component in
the $f_0(1500)$ \cite{closeamsler,closezhao,giacosa,hllz}, while others
concluded that this lightest glueball forms the primary component in the
$f_0(1710)$ \cite{lw99,ccl} and still others invoked important contributions
from $q \bar q q \bar q$ states \cite{schechter,maiani} (see also
\cite{narison}).  Further data and analyses should help to elucidate this
situation \cite{besreview,earlierreviews,bugg1370}.

\section{Gluon-Glueball Duality in Two Simplified Contexts}

To explain our notion of gluon-glueball duality, we begin with two simplified
forms of QCD, namely (i) withhout any light quarks, and (ii) in the large-$N_c$
limit.  Let us first consider the case of no light quarks. For definiteness, we
imagine the standard model with one generation of fermions with quarks $U$ and
$D$ having masses $m_U, \ m_D >> \Lambda_{QCD}$.  We denote these quarks
collectively as $Q$.  We next consider the favored reaction for glueball
production, namely the production, in $e^+e^-$ annihilation, of the
orthoquarkonium $Q \bar Q$ state, followed by its radiative decay $Q \bar Q \to
\gamma gg$.  An important feature of this world is that a number of the
lowest-lying glueball states would be stable.  Indeed, using the lattice
estimates of low-lying glueball masses listed above, all six of the states
listed would be stable; in order for a heavier glueball to be kinematically
allowed to decay to two of the lightest glueballs, it would necessarily have a
mass greater than about 3.4 GeV.  Thus, the invariant mass distribution
$dN/dm_G$ for the mass of the gluonic states recoiling against the photon in
the radiative orthoquarkonium decay $Q \bar Q \to \gamma gg$, i.e., at the
physical level, $Q \bar Q \to \gamma + X_{GB}$, where $X_{GB}$ denotes a
glueball, would exhibit very sharp resonances for $m_{X_{GB}}$ equal to the
mass of each of the stable glueballs, and then finite-width resonances for the
higher-lying unstable glueballs, up to the kinematic limit allowed by the mass
of the original orthoquarkonium state.  The statement of gluon-glueball duality
would be that, with appropriate smearing,
\beq
\int \, \left ( \frac{dN}{dm} \right )_{GB} \, dm = 
\int \, \left ( \frac{dN}{dm} \right )_{gg} \, dm \ , 
\label{ggdual}
\eeq
where the first integral is over physical glueball
final states and the second integral denotes the perturbative calculation of
$dN/dm$, where $m$ is the invariant mass of the $gg$ subsystem in the decay 
$Q \bar Q \to \gamma gg$. In terms of the overall $e^+e^-$ cross section, the
gluon-glueball duality would be the relation, with appropriate smearing, 
\beqs
& & \sum_{GB} d \sigma(e^+e^- \to n^3S_1(Q \bar Q) \to \gamma + X_{GB}) \cr\cr
& \simeq & d \sigma(e^+e^- \to n^3S_1(Q \bar Q) \to \gamma gg) \ , 
\label{ggdual2}
\eeqs
where again the second term represents the perturbative calculation of the
production and decay. 

   In the $N_c \to \infty$ limit, $q \bar q$ mesons and glueballs become
stable, as indicated by Eqs. (\ref{gammaqqbarlargenc}) and
(\ref{gammagblargenc}).  Furthermore, there is no mixing between glueballs and
$q \bar q$ mesons.  Here, gluon-glueball duality takes a particularly
simple form.  With $N_c$ large but finite, so as to allow for the radiative
decay of the heavy orthoquarkonium state, this duality would again be expressed
via the relations (\ref{ggdual}) and (\ref{ggdual2}).  Quark-hadron duality
also takes a particularly simple form in this large-$N_c$ QCD.  This type of
connection between sums over resonances and properties of the underlying quarks
and gluons was previously used with QCD sum rules to study correlators of
various operators \cite{svz}-\cite{narison}.

\section{Gluon-Glueball Duality in QCD}

  We next discuss our notion of gluon-glueball duality in real QCD.  An
important part of our discussion of this duality in the radiative decay of a
heavy orthoquarkonium state $Q \bar Q \to \gamma gg$ is a careful treatment of
the temporal evolution of the $gg$ subsystem, as it is initially produced, as
the gluons bind to form a proto-glueball, as this glueball mixes with a $q \bar
q$ component, and as it finally decays.  To understand gluon-glueball duality,
it is crucial to analyze the time evolution and hierarchy of time scales
relevant to the $Q \bar Q \to \gamma gg$ decay, as compared with the production
of mesons in a reaction such as $e^+e^- \to q \bar q$.  A general statement
concerns the time required for the formation of color-singlet states from the
respective initial $q \bar q$ and $gg$ states.  Given the fact that QCD
confines on a scale $\Lambda_{QCD}$ and that hadrons have a corresponding size
\beq
r_{had.} \simeq \frac{1}{m_\pi} \simeq 1 \ {\rm fm} \ , 
\label{rhad}
\eeq
and given the causality condition that information cannot be communicated any
faster than at the speed of light, it follows that a minimum time associated
with the formation of color-singlet hadronic states is
\beq
t_{had.} = \frac{r_{had.}}{c} \simeq 0.3 \times 10^{-23} \ {\rm sec}
\label{thad}
\eeq
(where we have explicitly indicated the speed of light, $c$).  This is a rough
estimate, accurate to a factor of order unity.  For example, given that a
glueball is represented by a closed string, one could consider a special case
in which the closed string forms a circle, and one might argue that it is the
circumference of this circle rather than the radius that is of order 1 fm.  In
this case, the radius would be $1/(2 \pi)$ fm and the time taken for the
formation, involving motion of the gluons outward from the center of the circle
would be smaller than the value given in Eq. (\ref{thad}) by the factor
$2\pi$.  Because of the asymptotic freedom of QCD, for both (i) $e^+e^- \to
\bar q q$ at center-of-mass energy $\sqrt{s} >> \Lambda_{QCD}$ and (ii) the
radiative decay of heavy orthoquarkonium $\bar Q Q \to \gamma gg$, there exists
a sufficiently short time $t_{pert.}$ such that for times $t < t_{pert.}$ the
physics can be described using perturbative QCD. This satisfies the inequality
\beq
t_{pert.} < \frac{1}{\Lambda_{QCD}} \sim t_{had.}  \ . 
\label{tpert}
\eeq
Given the precocious scaling behavior of QCD, it is not necessary that
$t_{pert.} << t_{had.}$.  For the two specifc cases under discussion, one could
take $t_{pert.} \sim 1/\sqrt{s}$ for the reaction $e^+e^- \to q \bar q$ and
$t_{pert.} \sim 1/(2m_Q)$ for the decay $Q \bar Q \to \gamma gg$.  A typical
value would be $t_{pert.} \sim 1/(3 \ {\rm GeV}) \simeq 10^{-25} \ {\rm sec}$.
Thus, in the reaction $e^+e^- \to \bar q q$, during the time interval $0 < t <
t_{pert.}$, the $q$ and $\bar q$ recede from each other in an approximately
perturbative manner, with the first modification being the emission of a gluon,
leading to a $q \bar q$ subsystem in a color octet state together with the
emitted gluon $g$.  In the radiative decay of the heavy orthoquarkonium state
$Q \bar Q \to \gamma gg$, during the time interval $0 < t < t_{pert.}$, the
$gg$ final-state subsystem mainly evolves into more gluons via $g \to gg$
splittings.  As noted above, this gluon splitting occurs at leading order in
the large-$N_c$ limit, in contrast to the $q \to q + g$ or $g \to (q \bar q)_8$
processes, which start to mix $q \bar q$ with the initially purely gluonic $gg$
subsystem.

   After a time $t_{MF}$, where $MF$ stands for ``meson formation'', the
initial $\bar q q$ system will bind to form a meson, and after a corresponding
time $t_{GBF}$, where $GBF$ stands for ``glueball formation'', the initial $gg$
system will bind to form a glueball.  From the causality argument above, one 
has the general inequalities
\beq
t_{MF}, \ t_{GBF} \ge t_{had.} 
\label{tfgen}
\eeq
and hence also the obvious inequalities $t_{MF}, \ t_{GBF} \ge t_{pert.}$. The
$q$ and $\bar q$ in the meson, and the gluons in the glueball have minimum
bound-state momenta $k_{min} \sim \Lambda_{QCD}$ because of confinement
\cite{lmax}.  Several factors are relevant for the hadronic formation times
$t_{MF}$ and $t_{GBF}$, including (i) the intrinsic QCD hadronization time
scale $t_{had.}$, (ii) the mixing of $q \bar q$ and gluonic states to form mass
eigenstates, (iii) the decay widths $\Gamma_i$ of various mesons and glueballs
and (iv) especially importantly for our current discussion, the density of
meson and glueball states, $n(m)_M$ and $n(m)_{GB}$. The quantum-mechanical
uncertainty relation $\Delta E \, \Delta t \gsim \hbar/2$ implies that the
observation time interval $\Delta t$ needed for an observer to measure the
spectrum of states with a resolution in mass $\Delta m$ is bounded below by
$\Delta t \ge (\hbar/2)/\Delta m$.  Here $\Delta m$ is set by a combination of
the density of states with the same quantum numbers (isospin and $J^{PC}$) and
by the widths of these states.  Let us consider a glueball search conducted in
the range of masses $m_{GB} = 1.5-3$ GeV.  Given the inequality in the density
of glueball versus $q \bar q$ meson states in Eq. (\ref{nratio}), it follows
that the time needed to experimentally measure and resolve glueball states is
shorter than that needed for $q \bar q$ meson states. Using the hadronic string
model as a theoretical guide, which is consistent with the observed states in
the relevant mass region, one has, roughly,
\beq
t_{GBF} \simeq \frac{t_{MF}}{4} \ . 
\label{ttrel}
\eeq

This leads us to suggest a different picture of glueball production than the
one that is often used in analyses of experimental data on glueball
searches. Conventional analyses use meson mass eigenstates that are linear
combinations of $q \bar q$ states and gluonic states.  Our new point is that it
is crucial to take into account the actual temporal formation of the glueball
states.  Given that the glueball formation time is shorter than the meson
formation time, with $t_{GBF} \simeq t_{MF}/4$ being a reasonable estimate, the
glueball forms before significant mixing with the $q \bar q$ sector takes
place.  A concrete realization of both the $\bar q q \leftrightarrow {\rm
meson}$ and $gg \leftrightarrow {\rm glueball}$ dualities can be obtained as
follows. Starting with an initial pure $\bar q q$ entrance state, we implement
the duality by letting the unitary QCD evolution operator, formally expressed
as $U(t)=e^{-iHt}$, operate on this state, where here $H$ denotes the QCD
Hamiltonian, yielding
\beq
U(t_{MF}) \, |\bar q q (t=0)\rangle = |M, \ {\rm meson} \rangle \ . 
\label{uopmeson}
\eeq
That is, the evolution over this time interval will yield a physical 
$\bar q q$ meson resonance. In a similar manner, in a purely gluonic sector 
\beq
U(t_{GBF}) \, |gg(t=0)\rangle = |GB, \ {\rm glueball}\rangle \ . 
\label{uopglueball}
\eeq
The two gluons in the $gg$ subsystem produced in the radiative orthoquarkonium
decay $\bar Q Q \to \gamma gg$ emerge from spacetime points that are
separated by a small distance $\Delta r \sim 1/m_Q$, where $Q=c$ or $b$ is a
heavy quark. This is not precisely the same as the production of a scalar
glueball by the action of the local operator
\beq
S(x) = G_{\mu\nu}(x)G^{\mu\nu}(x)
\label{s}
\eeq
on the vacuum.  However, a semiclassical argument leads to the
conclusion that the $gg$ usually bind with $L=0$ relative orbital angular
momentum.  For example, in the case $Q=b$, the spatial separation of the points
where the two gluons are emitted is $\Delta r \sim 1/m_b \sim 0.2 \ {\rm
GeV}^{-1}$. The 3-momenta of the gluons in the $gg$ rest frame are $|{\vec
k}_g| \sim m_{GB}/2 \sim O(1)$ GeV. The resultant average value of the relative
orbital angular momentum is
\beq
 \langle L \rangle \sim |{\vec k}_g|\Delta r \lsim 0.2 \ .
\label{angmom}
\eeq
Hence, one expects that this production mechanism will yield mainly glueball 
states with $L=0$, namely the $0^{++}$ and $2^{++}$ mentioned before. 

In the radiative $Q \bar Q \to \gamma gg$ decay of orthoquarkonium, the $gg$
subsystem is manifestly purely gluonic to start with, and mixing with $q \bar
q$ components occurs subsequently.  In the large-$N_c$ limit, this mixing is
suppressed by $1/N_c$, which has led to the common expectation that there could
be hadrons that are primarily gluonic, with only a small $q \bar q$ component.
Our estimate that $t_{GBF} \simeq t_M/4$, in conjunction with the suggestion
from large-$N_c$ arguments that mixing of gluonic and $q \bar q$ components may
be rather small, leads us to the important inference that the gluon-glueball
duality could hold reasonably well in full QCD as well as in the simplified
contexts which we initially used to introduce it.  It is understood that there
will be some corrections due to the the mixing of the gluons in a primarily
gluonic hadron with $q \bar q$ states.

Let us next consider the longer times required for the $\bar q q$ meson, or the
glueball, to decay into hadrons that are stable with respect to the strong
interactions.  The formation and decay times for the meson resonances are
comparable, although $\tau_M \gsim t_{MF} > t_{pert.}$, and similarly for the
glueballs, one has $\tau_{GB} \gsim t_{GBF} > t_{pert.}$.  This is to be
contrasted with the situation for a very heavy quark, namely the top quark,
which decays weakly before it can form color-singlet hadronic $t \bar t$ or $t
\bar q$ states.  To the extent that the large-$N_c$ limit is applicable to QCD,
one expects that, other factors such as phase space being equal, the lifetime
$\tau_{GB}$ might be somewhat longer than $\tau_M$, i.e., the glueball width
might be somewhat smaller than that for a $q \bar q$ meson of comparable mass.
However, in actual QCD, glueball widths may not be suppressed, and may, indeed,
be of order 100-300 MeV.  This would be somewhat analogous to the situation
with the $\eta'$ meson; in the $N_c \to \infty$ limit (with $\lambda \equiv g^2
N_c$ fixed), instanton effects are exponentially suppressed by the factor
$\exp(-8\pi^2 /g^2) = \exp(-8\pi^2 N_c/\lambda)$, so that U(1)$_A$ is a good
global symmetry and the isoscalar pseudoscalar meson $\eta'$ is an approximate
Nambu-Goldstone boson.  However, in real QCD the $\eta'$ is rather heavy, with
a mass of 958 MeV.  An important point is that, with the hierarchy of time
scales that we have noted, the glueball decays by popping two pairs of light $q
\bar q$ quarks out of the vacuum to produce the two final-state mesons
($\pi$'s, $K$'s, etc.).  This process is essentially equivalent to the process
by which the initially purely gluonic state acquires a $q \bar q$ component.

\section{Further Possible Insight from Lattice QCD}

Lattice calculations have the appeal of providing a fully nonperturbative tool
for studying the properties of QCD, and the advantage of being able to be
continually improved with the use of larger lattices, longer running times,
improved lattice actions, and careful analysis of statistical and systematic
uncertainties.  Most lattice QCD calculations of glueball masses have been
performed using the quenched approximation. Some unquenched calculations have
also been reported \cite{hartteper}. Both the necessity of evaluating the
fermion determinant and the related presence of disconnected flavor loops
appearing in unquenched calculations make these calculations more difficult
than computations in quenched QCD.  We suggest that it would be worthwhile for
lattice gauge simulations to address some of the issues that we have raised in
this paper.  We are interested not just in minor shifts of the glueball
spectrum, but rather in finding the time $t^*$ by which the admixture of the
initial glue state with the $\bar q q$ and multi-quark states becomes
significant. For this purpose it could be useful to study the correlator $C(t)
= \langle S(0) S(t)\rangle$ of the above-mentioned scalar glueball operator and
examine how its Euclidean time dependence might differ from a simple
exponential of the form $\exp(-m(0^{++}) \, t)$.  (Here, it is understood that
one would ideally have removed the effects of higher-lying glueball states with
the same $J^{PC}=0^{++}$ and also that one would have taken account of effects
due to periodic lattice boundary conditions.) For long, asymptotic times $t$
such that $t >> 1/(2m_\pi)$, the behavior of this scalar correlator $C(t)$ is
controlled by the lowest-mass $s$-channel threshold, namely that for the $2\pi$
final state, but we are interested in shorter times.  Similar calculations
could be performed for the $2^{++}$ glueball state by using an appropriate
color-singlet tensor correlator.  Assessing the full lifetime until the
glueball decays into final hadrons that are stable with respect to the strong
interactions is challenging, but is not essential for our purposes here.

\section{Application to Experimental Searches for Glueballs}

In this section we apply our notion of gluon-glueball duality to suggest a
method that could be useful in experimental searches for glueballs in radiative
orthoquarkonium decays, in particular, those involving the $\Upsilon(1S)$,
$\Upsilon(2S)$, and $\Upsilon(3S)$ states. There are very high-statistics data
sets from radiative $J/\psi$ decays, which have been used quite effectively for
glueball searches.  However, radiative $\Upsilon$ decays allow one to search in
a wider mass range and reduce phase space suppression for decays into final
states involving more massive glueballs. While a major purpose of the
experiments at BABAR and Belle was to study $B$ physics and CP violation, they
accumulated of order $10^9$ events from decays of $\Upsilon(1S)$,
$\Upsilon(2S)$, and $\Upsilon(3S)$, as well as the $\Upsilon(4S)$ state that
provided a copious source of $B_d$ mesons \cite{babar_res,belle_res}. These
data extended the already impressive data sets collected by the CLEO experiment
at CESR in its later years of high-intensity running
\cite{cleo3_upsilon_topipi,cleo3_etasearch}.  For an average radiative decay
branching ratio of 1.5 \% we expect of order $1.5 \times 10^7$ radiative decays
in the BABAR and Belle data. In the radiative $\Upsilon(nS) \to \gamma gg$
(with $n=1,2,3$) we label the four-momenta of the outgoing photon and gluons as
$k_\gamma$, $k_1$, and $k_2$, and recall that it is necessary to symmetrize the
amplitude under the interchange $k_1 \leftrightarrow k_2$ to take account of
the two identical bosons in the (perturbative) final state.  If one makes the
approximation, in the perturbative calculation of the amplitude, that the
outgoing (massless) gluons interact only very weakly with each other, it
follows that the three invariant mass combinations $(k_1+k_2)^2$,
$(k_\gamma+k_1)^2$, and $(k_\gamma + k_2)^2$ are uniformly distributed over the
Dalitz plot, which becomes an equilateral triangle.  In this Dalitz plot, the
region of interest, which is assumed here to be dominanted by the lowest-lying
glueballs, is then a rectangular strip adjacent to the bottom of the triangle.
The total area of this region is $2 \cdot (2.7)^2/100$, i.e. 15 \% of the total
area of the Dalitz plot and hence includes approximately $2 \times 10^6$
events. The notion of gluon-glueball duality that we have discussed then leads
us to the suggestion to analyze the inclusive mass distribution of these $2
\times 10^6$ events. This avoids any bias due to post-selection by the final
channel (which might prefer specific final $\bar q q$ or multiquark
resonances).  Our use of gluon-glueball duality is analogous to the use of
quark-hadron duality in the sense that both of these dualities relate inclusive
channels and sums of exclusive channels in the respective particle processes.
For notational simplicity, we denote $X \equiv X_{GB}$.  Clearly, only a crude
resolution $\Delta M_X \simeq 0.5$ GeV is needed to resolve the two
well-separated lowest-lying glueball states with $J^{PC}=0^{++}$ and $2^{++}$
(or the excited $0^{++}$ state). Let us denote the invariant mass squared of
the $gg$ subsystem as $M_X^2 = (k_1+k_2)^2$ and take particle energies to be
measured in the rest frame of the decaying $Q \bar Q$ orthoquarkonium
state. The elementary kinematic relation
\beq
M_{Q \bar Q}^2 = (k_\gamma + k_1 + k_2)^2 = 
2E_\gamma(M_{Q \bar Q} - E_\gamma) + M_X^2
\label{decaykinematics}
\eeq
implies that 
\beq
\Delta M_X = \frac{(2E_\gamma - M_{Q \bar Q})\Delta E_\gamma}{M_X} \ . 
\label{delta_mx}
\eeq
As an illustration, we consider the BABAR detector \cite{babar_res}; similar
numbers apply for the Belle detector \cite{belle_res}. The fractional
resolution $(\Delta E_\gamma)/E_\gamma$, of the measurement of the photon
energy by the electromagnetic calorimeter of this detector varies from from
about 2 to 3 \% over the range of $E_\gamma$ from $\sim 8$ GeV to 1 GeV
\cite{babar_res}.  Hence, the resultant resolution $\Delta M_X$ from
Eq. (\ref{delta_mx}), for the radiative decay of the $\Upsilon(1S)$, varies
from approximately 0.55 GeV to 0.27 GeV as $M_X$ varies from 1.7 GeV to 2.4
GeV.  For the radiative decay of the $\Upsilon(2S)$ the resolution $\Delta M_X$
varies from about 0.65 GeV to 0.33 GeV as $M_X$ varies from 1.7 GeV to 2.4 GeV.
Considering the very high statistics of the data sets obtained by BABAR and
Belle, this analysis could give useful information about glueballs via broad
deviations from the phase space distribution that would occur in their absence.
This analysis presumes that one takes careful account of pure quantum
electrodynamic (QED) backgrounds and corrections. By insisting on some hadronic
activity in the detector, one may reduce such QED backgrounds without excessive
biasing such as would result if one were to fully reconstruct the final
hadronic state.  Obviously, the experimental procedure sketched here in the
broadest terms is challenging.  Nevertheless, one has observed how much useful
new data BABAR and Belle have obtained concerning new hadronic states involving
charm quarks, including $X(3782)$, new $D_s$ states, and others. Provided that
our analysis of the time evolution of the glueball production process discssed
above is correct, then we believe that these facilities have the potential to
considerably clarify the lingering puzzles in glueball physics.

\section{Conclusions}

In this paper we have presented a different picture of glueball production than
the one commonly used in current analyses of data.  Using the hadronic string
model, we have given a quantitative estimate of the smaller density of states
of glueballs (closed strings) in the region of $\sim 2$ GeV, as compared with
$q \bar q$ mesons (open strings), and, from basic quantum mechanics, we have
inferred a resultant hierarchy of formation times of observable (resolvable)
glueballs, as compared with $q \bar q$ mesons, namely Eq. (\ref{ttrel}).  On
the basis of this, together with the suggestion from the large-$N_c$ expansion
that mixing between glueball and $q \bar q$ states may be suppressed, we have
argued that the glueballs produced in radiative orthoquarkonium decay could
plausibly form without substantial mixing with $q \bar q$ states.  This
motivates a notion of gluon-glueball duality, which we have presented, namely
that the summation over sufficiently many glueball states produced in radiative
orthoquarkonium decay $Q \bar Q \to \gamma gg$, appropriately smeared, could be
well fit with the perturbative calculation of this process.  We have applied
this notion of gluon-glueball duality to suggest a method that could be useful
in experimental searches for glueballs using radiative decays of the
$\Upsilon(1S)$, $\Upsilon(2S)$, and $\Upsilon(3S)$ states using the large data
sets that are currently available on these decays.

Acknowledgments: S.N. thanks A. Jawahery and R.S. thanks S. Brodsky for
helpful discussions.  The research of R. S. was partially supported by the
grant NSF-PHY-06-53342.

\vfill
\eject
\end{document}